\documentclass[letterpaper, 10 pt, conference]{ieeeconf}  % Comment this line out if you need a4paper

\IEEEoverridecommandlockouts                              % This command is only needed if 
\usepackage{cite}
\usepackage{amsmath,amsfonts,amssymb}
\usepackage{mathtools}
\usepackage{soul}
\usepackage{xcolor}
\usepackage{bm}
\usepackage{graphicx}
\usepackage{textcomp}
\usepackage{xcolor}
\usepackage{thmtools}
\usepackage{subcaption}
\usepackage{caption}
\usepackage{comment}
\usepackage{mdframed}
\usepackage{algorithm}
\usepackage{algpseudocode}
\usepackage{ulem}
\usepackage{float}
\usepackage{fancyhdr}

\newtheorem{definition}{Definition}
\newtheorem{theorem}{Theorem}

\title{\bf
	Learning Robust Regions of Attraction Using Rollout-Enhanced Physics-Informed Neural Networks with Policy Iteration
}

\author{Junkai Wang, Yuxuan Zhao, Mi Zhou and Fumin Zhang % <-this % stops a space
\thanks{Junkai Wang and Mi Zhou are with the School of Electrical and Computer Engineering, Georgia Institute of Technology, Atlanta, GA 30308, USA. Email: {\tt\small wangjk@gatech.edu, mzhou91@gatech.edu}.
Yuxuan Zhao and Fumin Zhang are with the Cheng Kar-Shun Robotics Institute, Hong Kong University of Science and Technology, Hong Kong, China. Email: {\tt\small yxz9711@gmail.com, eefumin@ust.hk}.\newline
Under review for publication. Copyright 2025 by the authors.
}
}

\begin{document}

\maketitle
\thispagestyle{empty}
\pagestyle{empty}

%%%%%%%%%%%%%%%%%%%%%%%%%%%%%%%%%%%%%%%%%%%%%%%%%%%%%%%%%%%%%%%%%%%%%%%%%%%%%%%%
\begin{abstract}
The region of attraction is a key metric of the robustness of systems. This paper addresses the numerical solution of the generalized Zubov’s equation, which produces a special Lyapunov function characterizing the robust region of attraction for perturbed systems. To handle the highly nonlinear characteristic of the generalized Zubov's equation, we propose a physics-informed neural network framework that employs a policy iteration training scheme with rollout to approximate the viscosity solution. In addition to computing the optimal disturbance during the policy improvement process, we incorporate neural network–generated value estimates as anchor points to facilitate the training procedure to prevent singularities in both low- and high-dimensional systems. Numerical simulations validate the effectiveness of the proposed approach.
\end{abstract}
%%%%%%%%%%%%%%%%%%%%%%%%%%%%%%%%%%%%%%%%%%%%%%%%%%%%%%%%%%%%%%%%%%%%%%%%%%%%%%%%

%
%%==== Introduction ====
%
\section{Introduction}
\label{sec:intro} 

% Cite more reference
The stability of dynamical systems is a fundamental research topic in control theory, and Lyapunov functions are the most widely used tool for certifying stability. In addition to qualitative analysis of stability, the region of attraction (ROA), within which all initial states are guaranteed to converge to the corresponding equilibrium, provides a quantitative measure of stability. Traditional approaches for constructing Lyapunov functions include sum of squares, linear programming, and linear matrix inequality \cite{Peter2015LFreview}. However, these methods impose strong structural requirements on the target system, such as linearity or polynomial dynamics. With the rise of machine learning, neural network–based methods have emerged as a powerful alternative for constructing Lyapunov functions. In \cite{long1993feedback, Abate2021LNN, gaby2022lyapunov, dai2021lyapunov}, the Lyapunov function itself is represented by a neural network, and its parameters are trained to satisfy Lyapunov conditions for general nonlinear systems by minimizing a loss function that penalizes violations of positive definiteness and the Lie derivative negativity condition.

Although Lyapunov functions provide certificates of system stability, the region of attraction estimated from the largest sublevel set of a Lyapunov function is often conservative.
To mitigate this conservativeness, several learning-based methods have been proposed to improve ROA estimation.
For example, \cite{richards2018lyapunov} trains a neural network Lyapunov function to align with the shape of the largest safe region in the state space and iteratively expand it, but this method does not ensure convergence to the true ROA or monotonic expansion. \cite{mehrjou2020neural} improves upon this by iteratively enlarging the largest stable sublevel set through forward simulation. Other than enlarging the safety region, a more direct approach is to use the maximal Lyapunov function \cite{zubov1961methods,vannelli1985maximal}, which exactly characterizes the ROA through its sublevel set. This Lyapunov function is constructed by an infinite integral along the system trajectory, making its analytical form intractable. Although directly evaluating this integral for all initial states is infeasible, the maximal Lyapunov function satisfies Zubov's partial differential equation, for which the physics-informed neural networks (PINNs) \cite{RAISSI2019PINN} offer a promising approach to obtain its solution. A PINN is a machine learning model that embeds the physical laws, expressed as partial differential equations, into its loss function to solve forward and inverse problems while respecting underlying physics. In the related works \cite{kang2023data, Wang2024ACPINN}, the authors design loss functions that simultaneously fit the neural network to finite simulation data and enforce Zubov’s equation. \cite{Zhou2024ELM} further relaxes the dependency on forward simulations and achieves accurate results by using a loss function consisting of only residual and boundary loss.

For systems with bounded disturbances, the region of attraction extends to the robust region of attraction (RROA), defined as the set of all initial states that converge to the equilibrium despite worst-case disturbances. Similar to the unperturbed case, the RROA can be characterized by the sublevel set of the generalized maximal Lyapunov function. This value function is the viscosity solution of the generalized Zubov's equation, a nonlinear PDE that is even more challenging to solve than its unperturbed counterpart. While finite difference methods \cite{camilli2007regularization} can approximate this solution, they are severely limited by the curse of dimensionality. In this context, PINNs combined with policy iteration (PINN-PI), where the neural network is fitted in the policy evaluation phase and the policy is updated in the policy improvement phase, have been proven to be efficient in research areas, such as optimal control and Hamilton-Jacobi reachability analysis \cite{abu2005nearly, Cheng2007Fixed, darbon2020overcoming, jiang2016using, sirignano2018dgm, Tassa2007Least, meng2024physics, rubies2016recursive}. 
% Among them, \cite{meng2024physics} proposes methods for solving nonlinear optimal control problems using neural network-based policy iteration techniques and corresponding convergence analysis. 

Although PINN–PI can address nonlinear PDEs, estimating the RROA for high-dimensional systems remains challenging. Neural network training may fail to produce robust results and can become trapped in local minima due to the absence of a data loss term, which plays a crucial role in PINNs. To mitigate this issue, \cite{meng2024physics} introduced a loss term that aligns the neural controller near the origin with the linear controller derived from the Lyapunov equation for the linearized system, thereby providing a reference signal that guides training toward stability. Building on this idea, but aiming to provide reference information beyond just the local neighborhood of the origin, we draw inspiration from rollout in reinforcement learning, where simulated trajectories are used to estimate the value function. Therefore, we propose an alternative PINN–PI framework that incorporates rollout to generate anchor values as supervised labels for estimating the viscosity solution of the generalized Zubov's equation. This integration enhances both the accuracy and stability of training.

Therefore, the main contributions of this paper  are summarized as follows:  
\begin{enumerate}
    \item \textbf{Physics-informed neural network framework:} A neural network-based method for computing the RROA value function for perturbed systems.
    \item \textbf{Policy iteration integration:} A policy iteration mechanism tailored to handle the nonlinearity of the supremum operator in the generalized Zubov’s equation.
    \item \textbf{Rollout-based data augmentation:} Forward simulation is used to generate supervised anchor points, enhancing convergence and mitigating singularities.
\end{enumerate}

The remainder of the paper is organized as follows. In Section~\ref {sec:problem}, we introduce the problem formulation. In Section~\ref{sec:zubovmethod}, we present the derivation of the generalized Zubov equation. In Section~\ref{sec:NN}, a neural network-based method for RROA is presented. In Section~\ref{sec:experiments}, we conduct simulations to verify the proposed method. Section~\ref{sec:conclusions} concludes this paper.

%
%%====Problem Setup====
%
\section{Problem Formulation} 
\label{sec:problem}

Consider a perturbed system
\begin{align}\label{eqn:perturbedSystem}
\dot{\bm{x}}(t) = \bm{f}(\bm{x}(t), \bm{\delta}(t)),
\end{align}
where $\bm{x}(t)\in \mathbb{R}^n$ is the system state, $\bm{\delta}(t) \in \Delta\subset \mathbb{R}^m$ is a bounded time-varying disturbance and $\bm{f}:\mathbb{R}^n\times\mathbb{R}^m\rightarrow\mathbb{R}^n$ is a Lipschitz-continuous system function. Furthermore, the equilibrium is assumed to be at the origin and singular, i.e., $\bm{f}(\bm{0}, \bm{\delta}) = \bm{0},\ \forall \bm{\delta} \in \Delta$. 
Given an initial state $\bm{x}$ and disturbance $\bm{\delta}(\tau)$ defined over $\tau\in[0,t]$, the Lipschitz continuity of $\bm{f}$ ensures the existence and uniqueness of the corresponding trajectory, denoted by $\bm{\phi}(t; \bm{x}, \bm{\delta})$.
The definition of a robust region of attraction is given below.
\begin{definition}
    For a perturbed system (\ref{eqn:perturbedSystem}), the robust region of attraction is given by
    \begin{align*}
    \mathcal{A} = \{\bm{x}: \bm{\phi}(t;\bm{x},\bm{\delta}) \rightarrow \bm{0}\ \text{as} \ t\rightarrow +\infty,\; \forall \bm{\delta}\in \Delta\}.
    \end{align*}
\end{definition}

In this work, we propose a neural network-based method to identify RROAs for perturbed systems, under the assumption that such regions exist and are bounded.

\section{Generalized Zubov's Method for Robust Region of Attraction Estimation} \label{sec:zubovmethod}
In this section, we first introduce the background of Zubov's theorem for the region of attraction estimation and then introduce its extension to perturbed systems.

\subsection{Maximal Lyapunov Function and Zubov's Theorem}
Consider an autonomous system:
\begin{equation}\label{eqn:unperturbedSystem}
\begin{aligned}
    \dot{\bm{x}}(t) = \bm{f}(\bm{x}(t)),
\end{aligned}
\end{equation}
with an asymptotically stable equilibrium at $\bm{x} = \bm{0}$. To identify its region of attraction, a special type of Lyapunov function, called the maximal Lyapunov function and denoted by $V(\bm{x})$, can be constructed as proposed in \cite{vannelli1985maximal}. Its key properties are summarized below.
\begin{theorem}
Suppose there exists a set $\mathcal{A}\subseteq \mathbb{R}^n$ containing the origin in its interior, a continuously differentiable function $V:\mathcal{A} \rightarrow \mathbb{R}_+$, and a positive definite function $\Phi:\mathbb{R}^n\rightarrow \mathbb{R}_+$ satisfying the following conditions:
\begin{align*}
    &V(\bm{0}) = 0,\ V(\bm{x})>0,\; \forall \bm{x}\in\mathcal{A}\setminus\{\bm{0}\} ,\\
    &V(\bm{x})\rightarrow \infty\ \mathrm{as}\ \bm{x}\rightarrow \partial\mathcal{A}\ \mathrm{or}\ \|\bm{x} \|\rightarrow\infty,\\
    &\nabla V(\bm{x})^\intercal {f}(\bm{x}) = -\Phi(\bm{x}), \; \forall \bm{x}\in \mathcal{A}.
\end{align*}
Then $\mathcal{A}$ is the region of attraction of the origin.
\end{theorem}

A commonly employed construction of the maximal Lyapunov function by choosing $\Phi: = \|\cdot\|^2$ is given by 
% \begin{equation}\label{eq:maxLFDef}
%  \begin{aligned}
%     V(\bm{x}) = \left\{\begin{array}{ll}
% \int_0^{+\infty}\|\bm{\phi}(t;\bm{x})\|^2{\rm d}t     & \mathrm{if\ convergent},  \\
%     \infty     & \mathrm{otherwise}.
%     \end{array}\right.
% \end{aligned}   
% \end{equation}

\begin{equation}\label{eq:maxLFDef}
 \begin{aligned}
    V(\bm{x}) = 
\int_0^{+\infty}\|\bm{\phi}(t;\bm{x})\|^2{\rm d}t,   
\end{aligned}   
\end{equation}
where $\bm{\phi}(t;x)$ is the trajectory starting from $\bm{x}$ governed by Eqn. \eqref{eqn:unperturbedSystem}.

Although the maximal Lyapunov function characterizes the region of attraction through its finite domain, its unbounded nature near the boundary and at infinity impedes numerical approximation. Hence, Zubov's method addresses this by constructing a bounded counterpart.

\begin{theorem}[Zubov's Theorem]
Suppose there exists a set $\mathcal{A}\subseteq \mathbb{R}^n$ containing the origin in its interior, a continuously differentiable function $v:\mathcal{A} \rightarrow \mathbb{R}_+$, and a positive definite function $\Psi:\mathbb{R}^n\rightarrow \mathbb{R}$ satisfying the following conditions: 
% {\color{blue}maybe use $\bm{x}^*$ to replace $\bm{0}$}
\begin{align}
&v(\bm{0}) = 0,\ 0<v(\bm{x})<1, \; \forall \bm{x}\in\mathcal{A}\setminus\{\bm{0}\}, \nonumber\\ 
&v(\bm{x})\rightarrow 1\ \mathrm{as}\ \bm{x}\rightarrow \partial\mathcal{A}\ \mathrm{or}\ \|\bm{x}\|\rightarrow\infty,\nonumber\\
% &\Psi(\bm{0})=0,\ \Psi(\bm{x})>0\ \forall \bm{x}\in\mathcal{A}\setminus\{\bm{0}\},\\
&\nabla v(\bm{x})^\intercal f(\bm{x}) = -\Psi(\bm{x})(1-v(\bm{x})),\ \forall \bm{x}\in\mathcal{A}.\label{eq:derivative}
\end{align}
Then $\mathcal{A}$ is the region of attraction with respect to the asymptotically stable equilibrium origin.
\end{theorem}

A common transformation between the maximal Lyapunov function and its bounded counterpart is:
\begin{equation}\label{eq:transformation}
\begin{aligned}
    v(\bm{x}) = 1 - e^{-\alpha V(\bm{x})},\ \alpha>0.
\end{aligned}
\end{equation}
It is worth noting that $\alpha$ controls the steepness of the function $v(\bm{x})$, which has a strong impact on estimations. A steeper $v(\bm{x})$ tends to yield more conservative estimates and increase sensitivity to training errors.

% I have one thoughts here. through invertable NN, could we construct a NN + mapping function so that we could directly obtain the Lyapunov function
\subsection{Generalized Zubov's Theorem}
Now we introduce the generalization of Zubov's theorem to the perturbed system according to \cite{camilli2001generalization}. Consider the following non-negative value function $V:\mathbb{R}^n\rightarrow\mathbb{R}_+$, which is analogous to the maximal Lyapunov function in the unperturbed setting:

\begin{align}
    V(\bm{x}) = \sup_{\bm{\delta}\in\Delta} \int_{0}^{+\infty}g(\bm{\phi}(t;\bm{x},\bm{\delta}),\bm{\delta}(t))\mathrm{d}t,
\end{align}
where $g:\mathbb{R}^n\times\mathbb{R}^m\rightarrow\mathbb{R}_+$ is a nonnegative and Lipschitz continuous function.
A typical choice of $V(\bm{x})$ is:
\begin{align}\label{eq:generalizedMLF}
    V(\bm{x}) = \sup_{\bm{\delta} \in \Delta}\int_{0}^{+\infty}\|\bm{\phi}(t;\bm{x},\bm{\delta})\|^2dt,
\end{align}
which has the following properties:
\begin{enumerate}
    \item $V(\bm{x})<+\infty$ if and only if $\bm{x} \in \mathcal{A}$. 
    \item $V(\bm{0}) = 0$.
    \item $V(\bm{x})\rightarrow+\infty$ as $\bm{x}\rightarrow\partial\mathcal{A}$ or $\|\bm{x}\|\rightarrow\infty$.
\end{enumerate}

To obtain a bounded representation, we again apply the transformation Eqn. \eqref{eq:transformation} and acquire the generalized Zubov's Theorem.

\begin{theorem}[Generalized Zubov's Theorem]\label{thm:generalizedZubov}
Suppose there exists a set $\mathcal{A}\subseteq \mathbb{R}^n$ containing the origin in its interior, $v$ satisfies the following conditions:
\begin{align}
&v(\bm{0}) = 0,\ 0<v(\bm{x})<1, \; \forall \bm{x}\in\mathcal{A}\setminus\{\bm{0}\}, \nonumber\\ 
&v(\bm{x})\rightarrow 1\ \mathrm{as}\ \bm{x}\rightarrow \partial\mathcal{A}\ \mathrm{or}\  \|\bm{x}\|\rightarrow \infty, \nonumber
\end{align}
and is a viscosity solution of
\begin{align}\label{eqn:viscosity}
\sup_{\bm{\delta}\in\Delta}\left\{ \nabla v(\bm{x}) \bm{f}(\bm{x},\bm{\delta}) + \alpha(1 - v(\bm{x}))g(\bm{x}, \bm{\delta})\right\} = 0,\ x\in\mathbb{R}^n.
\end{align}
Then $\mathcal{A} = \{\bm{x}:v(\bm{x})<1\}$ is the RROA.
\end{theorem}

In general, it is difficult to get the viscosity solution of the generalized Zubov's equation analytically or to directly simulate the trajectory to get the value function because of the nonlinear nature of Eqn. \eqref{eqn:viscosity}. Existing tools \cite{camilli2007regularization} employ the gridding technique to numerically approximate the solution using dynamic programming. However, gridding techniques suffer from the curse of dimensionality, making exact solutions impractical.
Instead, we seek neural network-based approaches that offer a promising way to overcome this challenge.

\section{Neural Network Training for the Viscosity Solution of the Generalized Zubov's Equation} \label{sec:NN}
In this section, we discuss the framework of the physics-informed neural network with policy iteration to numerically compute the viscosity solution of the generalized Zubov's PDE.

\subsection{Neural Network Structure}
We use multilayer feedforward neural networks to approximate the value function $v(\bm{x})$. The feedforward neural network $v_{\bm{\theta}}:\mathbb{R}^n\rightarrow\mathbb{R}$ is expressed as follows:
\begin{equation}
\begin{aligned}
   v_{\bm{\theta}}(\bm{x}) = g_L\circ \bm{g}_{L-1}\circ \cdots \circ \bm{g}_{1}(\bm{x}), 
\end{aligned}
\end{equation}
where
\begin{equation}
\begin{aligned}
    &\bm{g}_{l}(\bm{y}) = \bm{\sigma}(\bm{W}_{l}\bm{y} + \bm{b}_l),\ \forall l\in \{1,\cdots,L-1\},\\
    &g_L(\bm{y}) = \bm{W}_L\bm{y}+b_L.
\end{aligned}
\end{equation}
Here $\bm{W}_l$ and $\bm{b}_l$ denote the weight matrices and bias vectors for layer $l$, which are all trainable parameters and denoted by $\bm{\theta}= \{\bm{W}_l, \bm{b}_l\}^{L}_{l=1}$. All hidden layers are assigned the same width $w$. $\bm{\sigma}$ is a vector-valued nonlinear activation function.
Although the Rectified Linear Unit (ReLU) is a widely utilized activation function in deep learning, its non-smooth nature can impede accurate gradient estimation, which is essential for our method, as elaborated in the subsequent section on loss function design. Therefore, we employ the sigmoid function as the activation function. It is noteworthy that the output layer is implemented as a linear layer to regularize excessively large parameter values in the hidden layers, which may result from the non-smooth boundary of the RROA.

\subsection{Rollout-Enhanced Policy Iteration Framework}
In this subsection, we introduce the policy iteration framework for training the physics-informed neural network to overcome the nonlinearity of the generalized Zubov's equation, and we employ rollout to assist the training.

Policy iteration is a classical algorithm in optimal control and reinforcement learning that alternates between two main steps:
1) \textbf{Policy evaluation}: estimate the value function under the current control policy.
2) \textbf{Policy improvement}: update the policy to reduce cost, typically by minimizing the Hamiltonian.
These two steps are repeated until convergence. 

However, in practical applications, relying solely on the PDE residual loss and boundary loss for training is likely to trap the optimization in local minima, leading to degenerate solutions such as the singularity case reported in \cite{camilli2007regularization}, in both low- and high-dimensional systems.

Our method differs from the previous PINN-PI methods in that $v_{\bm{\theta}}$ is also used to simulate forward trajectories under the estimated optimal disturbance. In PINN training, the data loss plays an important role in matching the neural network to the value function. However, the supremum operator in the generalized maximal Lyapunov function places difficulties in directly simulating and estimating the corresponding values. Although nonlinear model predictive control \cite{allgower2012nonlinear} can in principle solve this optimization problem, it is computationally expensive in both time and memory. Instead, we adopt an alternative approach by leveraging the current neural network to determine optimal disturbances along the trajectory to enable forward simulation, which shares the same idea as the rollout in reinforcement learning. These simulated trajectories are used to compute anchor values of the value function, which serve as supervised labels in the loss function. This step allows us to effectively train the PINN even in the absence of exact solution data. To sum up, the policy iteration is implemented as follows:

In the policy evaluation, the neural network parameters $\bm{\theta}$ are updated under given optimal disturbances $\bm{\delta^*}$ and the estimated anchor values $\hat{v}$ by gradient descent with the empirical loss.

In the policy improvement, we compute the optimal disturbance $\bm{\delta}^*$ by Eqn.~\eqref{eqn:viscosity} and forward simulate $\hat{v}$ for the next iteration.

The overall algorithm is detailed in Algorithm \ref{alg:PINN}.
\begin{algorithm}[h] 
\caption{Training Procedure of the Rollout-Enhanced PINN with Policy Iteration}
\label{alg:PINN}
\begin{algorithmic}[1]
\State \textbf{Input:} Dynamics function $\bm{f}(\bm{x}, \bm{\delta})$, sampling region $\Omega$, neural network $v_{\bm{\theta}}(\bm{x})$
\State \textbf{Output:} Trained neural network parameters $\bm{\theta}$

% Add Training Loop or Next Steps Here...
\State \textbf{Iteration Step:}
\For{each iteration $k = 1$ to $T$}
    \State \textbf{Policy Evaluation:}
        \State \hspace{1em} Get new neural network parameter $\bm{\theta}$ by
        \State \hspace{1em} \textbf{Repeat}
        
            \State\hspace{2em} Run gradient descent on $\bm{\theta}$ with $\hat{l}(\bm{\theta})$
    \State \hspace{1em} \textbf{Until} desired accuracy or max epochs reached

    \State \textbf{Policy Improvement:}
        \State \hspace{1em} Update optimal disturbance $\bm{\delta}^*$ and anchor values $\hat{v}$ for the next iteration 
\EndFor
\end{algorithmic}
\end{algorithm}

\subsection{Policy Evaluation Process}
The neural network training in the policy evaluation phase refers to the search for specific parameters $\bm{\theta}$ such that the empirical loss $\hat{l}(\bm{\theta})$ is minimized. To enable training of the neural network, we begin by sampling several sets of system states from a compact region $\Omega\subset\mathbb{R}^n$ that is assumed to contain the true RROA. These sampled states allow us to enforce boundary conditions, penalize the residuals of the PDE constraints, and guide the network toward matching the estimated value function across representative subsets of the state space. Based on these samples, we define the empirical loss function as follows:
\begin{align}\label{eqn:lossfunction}
    \hat{l}(\bm{\theta}) = \hat{l}_{\mathrm{boundary}}(\bm{\theta}) + \lambda_r\hat{l}_{\mathrm{residual}}(\bm{\theta}) + \lambda_d\hat{l}_{\mathrm{data}}(\bm{\theta}),
\end{align}
where $\lambda_r$ and $\lambda_d$ are hyperparameters.

We detail these three components as follows:
\begin{enumerate}
    \item 
$\hat{l}_{\mathrm{boundary}}({\bm{\theta}})$ penalizes the violation of the boundary conditions for both origin and states outside the RROA, i.e., $v_{\bm{\theta}}(\bm{0}) = 0$ and $v_{\bm{\theta}}(\bm{x}) = 1,\ \forall \bm{x}\in \mathcal{A}^c$. 
However, since the exact RROA is not known a priori, the latter condition cannot be directly enforced. 
Instead, we leverage the assumption that the sampling region $\Omega$ contains the RROA, which implies that any state on the boundary $\partial \Omega$ lies outside the RROA. Therefore, we enforce the alternative condition: $v_{\bm{\theta}}(\bm{x}) = 1,\ \forall \bm{x}\in\partial\Omega$, and define $\hat{l}_{\mathrm{boundary}}({\bm{\theta}})$ as:
\begin{equation}
\begin{aligned}
    \hat{l}_{\mathrm{boundary}}(\bm{\theta})= \left[ v_{\bm{\theta}}(\bm{0})\right]^2 +\frac{1}{M_b}\sum_{i=1}^{M_b}\left[v_{\bm{\theta}}(\bm{x}_i) - 1\right]^2,
\end{aligned}
\end{equation}
where $\{\bm{x}_i\}_{i=1}^{M_b}$ are states uniformly sampled on $\partial\Omega$.
\item $\hat{l}_{\mathrm{residual}}(\bm{\theta})$ penalizes the residual of Eqn. \eqref{eqn:viscosity}:
\begin{equation}
\begin{aligned}
\hat{l}_{\mathrm{residual}}(\bm{\bm{\theta}})=\frac{1}{M_r}\sum_{i=1}^{M_r}\left[ \nabla v_{\bm{\theta}}(\bm{x}_i)^\top\bm{f}(\bm{x}_i, \bm{\delta}^*_i)\right. \\
    \left.+\alpha(1-v_{\bm{\theta}}(\bm{x}_i))g(\bm{x}_i,\bm{\delta}^*_i)\right]^2,
\end{aligned}
\end{equation}
where $\{\bm{x}_i\}_{i=1}^{M_r}$ are states uniformly sampled in the interior of $\Omega$, and $\{\bm{\delta}^*_i\}_{i=1}^{M_r}$ are corresponding optimal disturbances which will be introduced in the subsection \ref{subsec:PI}.   
\item $\hat{l}_{\mathrm{data}}(\bm{\theta})$ enforces the neural network to approach the estimated value function.
\begin{equation}
\begin{aligned}
\hat{l}_{\mathrm{data}}(\bm{\theta}) = \frac{1}{M_d}\sum_{i=1}^{M_d}[v_{\bm{\theta}}(\bm{x}_i)-\hat{v}_i]^2,
\end{aligned}
\end{equation}
where $\{\bm{x}_i\}_{i=1}^{M_d}$ are states uniformly sampled in the interior of $\Omega$, and $\{\hat{v}_{i}\}_{i=1}^{M_d}$  are corresponding estimated values by forward simulation which will be introduced in the subsection \ref{subsec:PI}. 
\end{enumerate}
\subsection{Policy Improvement Process} \label{subsec:PI}
In this phase, we need to compute $\{\bm{\delta}^*_i\}_{i=1}^{M_r}$ and $\{\hat{v}_{i}\}_{i=1}^{M_d}$ according to the current neural network $v_{\bm{\theta}}(\bm{x})$.

The optimal disturbance $\bm{\delta_i^*}$ for each state point $\bm{x}_i$ is computed following the classical policy improvement strategy:
\begin{equation}
\begin{aligned}
    \bm{\delta}^*_{i} = \arg\sup_{\bm{\delta}\in\Delta}\left[ \nabla v_{\bm{\theta}}(\bm{x}_i)^\top\bm{f}(\bm{x}_{i}, \bm{\delta})\right. \\
    \left.+\alpha(1-v_{\bm{\theta}}(\bm{x}_i))g(\bm{x}_i,\bm{\delta})\right].
\end{aligned}\label{eq:optimalDisturbance}
\end{equation}

For $\{\hat{v}_{i}\}_{i=1}^{M_d}$, we first compute $\{\hat{V}_{i}\}_{i=1}^{M_d}$, and then deploy the transformation as in Eqn. \eqref{eq:transformation}. $\{\hat{V}_{i}\}_{i=1}^{M_d}$ is acquired by simulating a finite-time discrete Eqn. \eqref{eq:generalizedMLF}:
\begin{equation}
\begin{aligned}
    &\hat{V}_{i} = \sup_{\bm{\delta}\in\Delta}\sum_{k=0}^{K}\|\bm{x}_{i,k}\|^2\Delta t,\\
    \mathrm{s.t.}\quad
    &\bm{x}_{i,k+1} = \bm{x}_{i,k} + \bm{f}(\bm{x}_{i,k},\bm{\delta})\Delta t,\\
    &\bm{x}_{i,0} = \bm{x}_i,
\end{aligned}
\end{equation}
where $K$ denotes the total number of discrete time steps, $\bm{x}_{i,k}$ is the state of the system at the $k$-th discrete time step along a trajectory initialized at $\bm{x}_{i,0} = \bm{x}_{i}$, and $\Delta t$ is the fixed time step size. To handle the supremum operator in the forward simulation, we adopt a rollout that iteratively computes the optimal disturbance using Eqn.~\eqref{eq:optimalDisturbance} and then propagates the system trajectories. The key steps for computing $\{\hat{V}_{i}\}_{i=1}^{M_d}$ are as follows:
\begin{equation}
\begin{aligned}
    &\hat{V}_{i} = \sum_{k=0}^{K}\|\bm{x}_{i,k}\|^2\Delta t,\\
    \mathrm{s.t.}\quad
    &\bm{x}_{i,k+1} = \bm{x}_{i,k} + \bm{f}(\bm{x}_{i,k},\bm{\delta}_{i,k}^*)\Delta t,\\
    &\bm{\delta}^*_{i,k} =\arg\sup_{\bm{\delta}\in\Delta}\left[ \nabla v_{\bm{\theta}}(\bm{x}_{i,k})^\top\bm{f}(\bm{x}_{i,k}, \bm{\delta})\right. \\
    &\qquad \left.+\alpha(1-v_{\bm{\theta}}(\bm{x}_{i,k}))g(\bm{x}_{i,k},\bm{\delta})\right],\\
    &\bm{x}_{i,0} = \bm{x}_i,
\end{aligned}
\end{equation}
where $\bm{\delta}_{i,k}^*$ is the optimal disturbance at $\bm{x}_{i,k}$ based on current $v_{\bm{\theta}}(\bm{x})$.

\section{Simulation} \label{sec:experiments}
\subsection{Experiment Setting}
We demonstrate the effectiveness of the proposed method through numerical experiments in this section. All simulations were conducted on a PC equipped with an Intel Core i7-13700KF CPU and an NVIDIA RTX 4080 GPU, with 32 GB of RAM. The code was implemented in Python 3.9 using PyTorch, and all experiments were run on Windows 11 OS.

All neural network models were initialized using random seeds. Training was performed using Adam optimizer with a learning rate of 0.001. For the transformation equation Eqn. \eqref{eq:transformation}, we chose $\alpha = 0.5$ for all simulations. Other training parameters are listed in the Table. \ref{tab:parameters}.

\begin{table}[ht]
\centering
\caption{Training Parameters for Simulations}
\begin{tabular}{cccc}
\hline
\textbf{Parameters} & \textbf{Van der Pol} 
&\textbf{Inverted pendulum} & \textbf{10-d System} \\
\hline
$w$ & 50 &50 & 50\\
$L$ & 5 &5 & 5\\
$M_b$ & 20000 &20000 & 50000 \\
$M_r$ & 20000 &20000 & 50000 \\
$M_d$ & 2000 &2000 & 30000 \\
$\lambda_r$ &1 &1 &1\\
$\lambda_d$ &10 &10 &10\\
\hline
\end{tabular}
\label{tab:parameters}
\end{table}

\subsection{Experimental Results}
\subsubsection{Van der Pol system}
\begin{figure}[h]
    \centering
    \begin{subfigure}[t]{0.45\linewidth}        \includegraphics[width=\linewidth]{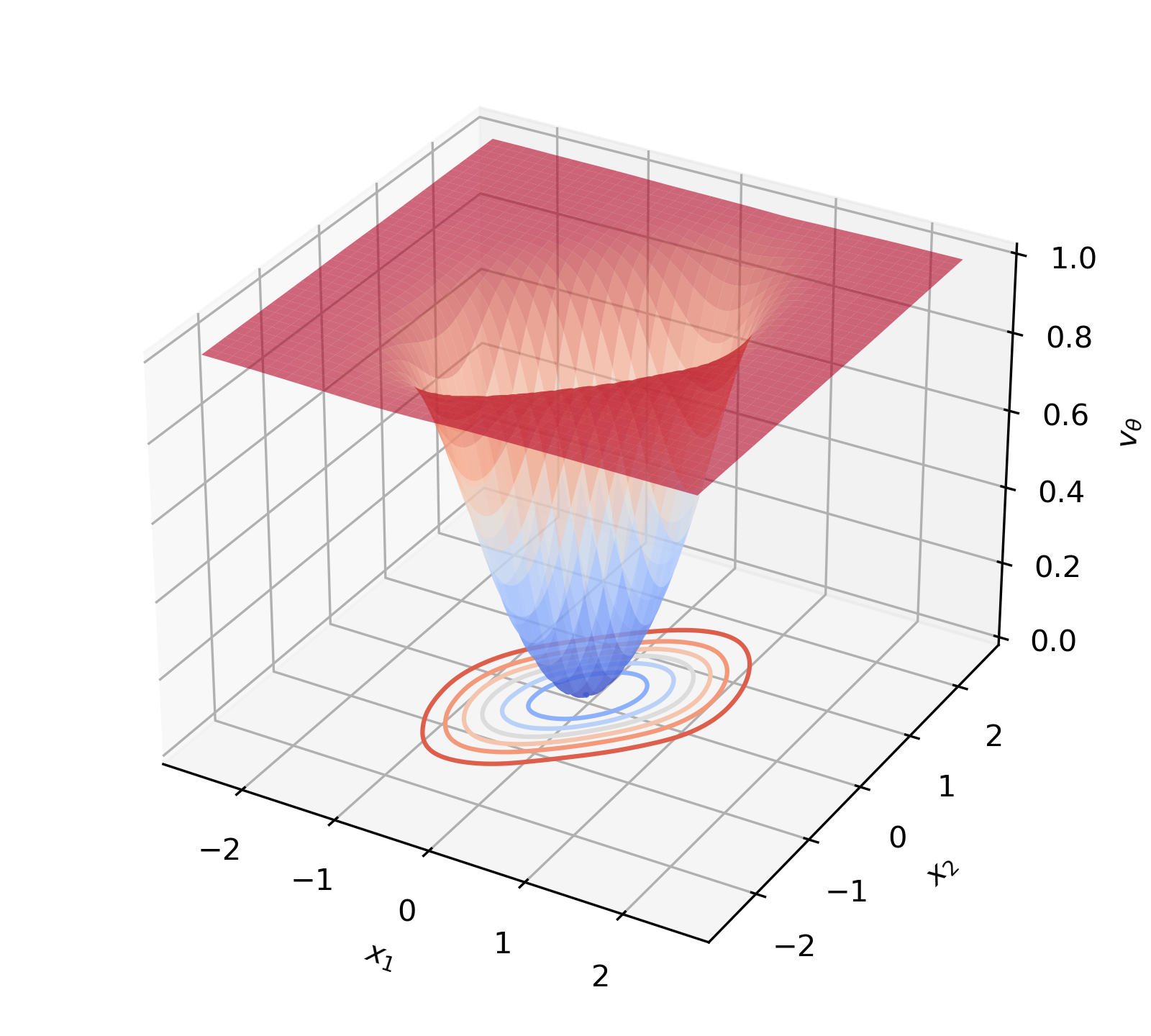}
        \caption{Learned $v_{\bm{\theta}}(\bm{x})$ of the perturbed Van der Pol system.}
        \label{subfig:VDPValueFunction}
    \end{subfigure}
    \hfill
    \begin{subfigure}[t]
    {0.45\linewidth}
    \includegraphics[width=\linewidth]{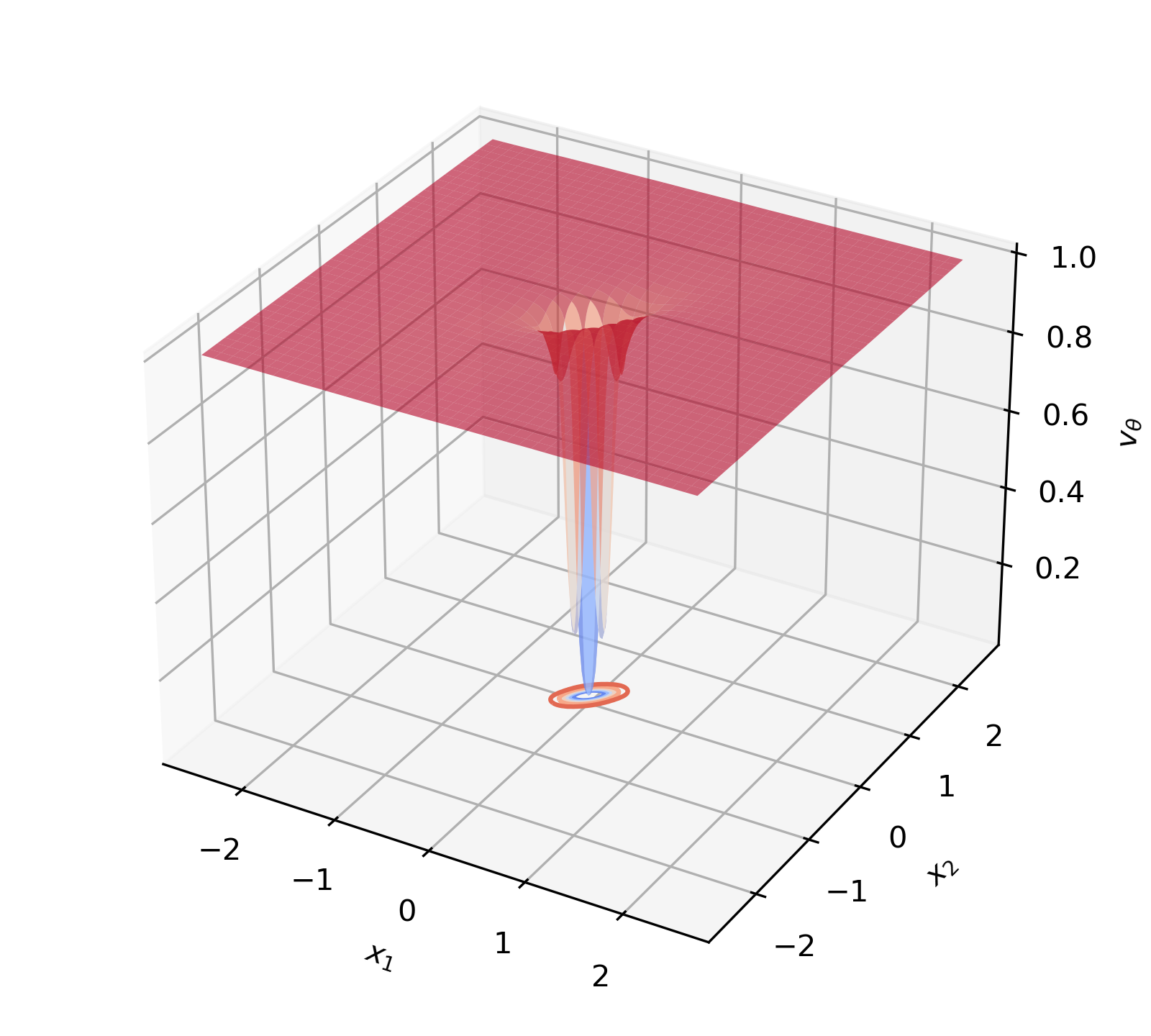}
    \caption{Learned $v_{\bm{\theta}}(\bm{x})$ of the perturbed Van der Pol system without rollout.}
    \label{subfig:VDPSingular}
    \end{subfigure}
    \hfill
    \begin{subfigure}[t]{0.45\linewidth}
    \includegraphics[width=\linewidth]{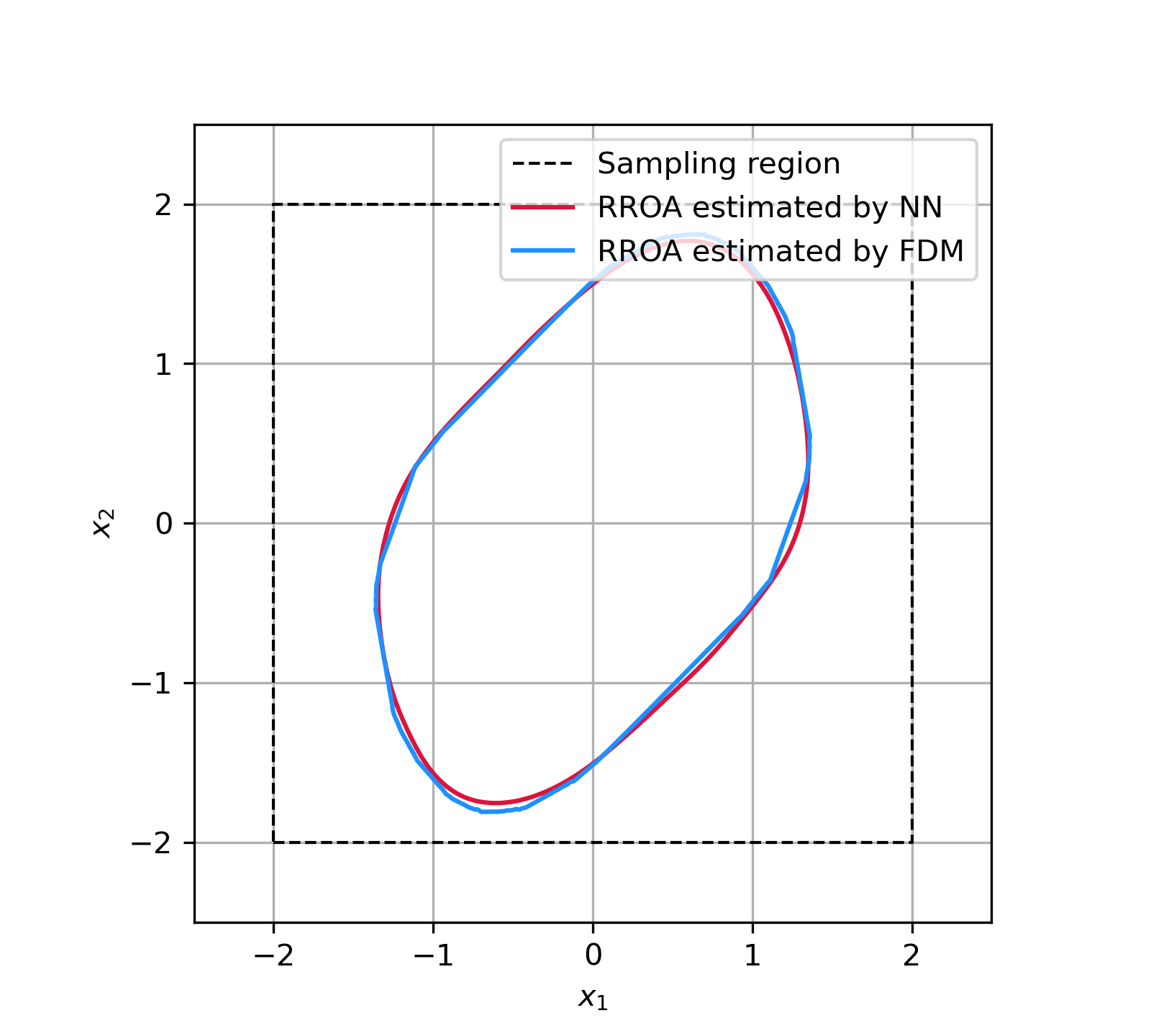}
        \caption{Learned RROA of the Van der Pol system.}
    \label{subfig:VDPRROA}
    \end{subfigure}
    \caption{The simulation result for the perturbed Van der Pol system.}
    \label{fig:VDPSystem}
\end{figure}

The first simulation was conducted on the well-known benchmark nonlinear system, the Van Der Pol system, augmented with perturbation terms:
\begin{equation}\label{eqn:VDP}
\begin{aligned}
    \dot{x}_1 &= -x_2 + \delta_1 x_1\\
    \dot{x}_2 &= x_1 - (1 - x_1^2)x_2 +\delta_2 x_2 ,
\end{aligned}
\end{equation}
where $(\delta_1, \delta_2)\in [-0.3, 0.3] \times[-0.1, 0.1]$. The sampling region was chosen as $\Omega = [-3,3]\times[-3, 3]$.
We first illustrate the necessity of rollout in training based on the results in Fig.\ref{fig:VDPSystem}. Fig.\ref{subfig:VDPSingular} shows the outcome when the empirical loss excludes $\hat{l}_{\mathrm{data}}(\bm{\theta})$, i.e., $\lambda_d=0$. The learned value function equals 1 almost everywhere on the sampling region except the small neighborhood around the origin, which is very similar to the singular case in \cite{camilli2007regularization}. The training loss remains low because the PDE is only violated in the vicinity of the origin. Elsewhere, the solution is flat (zero gradient) with a value of 1, which means the residual loss is disabled. 
To validate the accuracy of the proposed method, we compared it with the finite difference method (FDM). As shown in Fig.\ref{subfig:VDPRROA}, the red and blue contours represent the RROA estimated by the neural network-based method and the finite difference method, respectively. The FDM serves as the reference solution for validation. The simulation results demonstrate that the neural network-based method can accurately estimate the RROA.

\subsubsection{Inverted pendulum system}
\begin{figure}[h]
    \centering
    \begin{subfigure}[t]{0.45\linewidth}        \includegraphics[width=\linewidth]{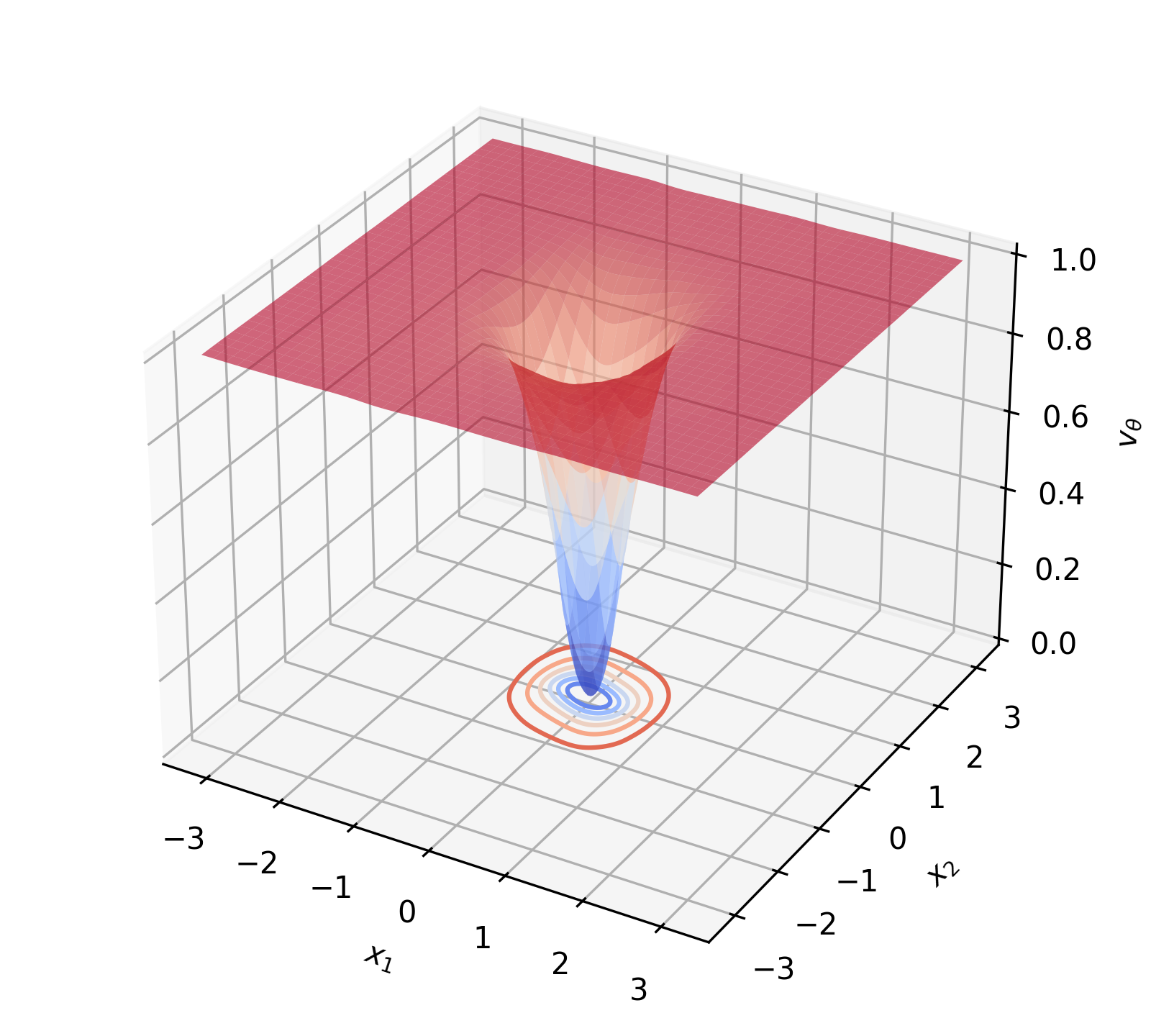}
        \caption{Learned $v_{\bm{\theta}}(\bm{x})$ of the perturbed inverted pendulum system.}
        \label{subfig:IPValueFunction}
    \end{subfigure}
    \hfill
    \begin{subfigure}[t]{0.45\linewidth}
    \includegraphics[width=\linewidth]{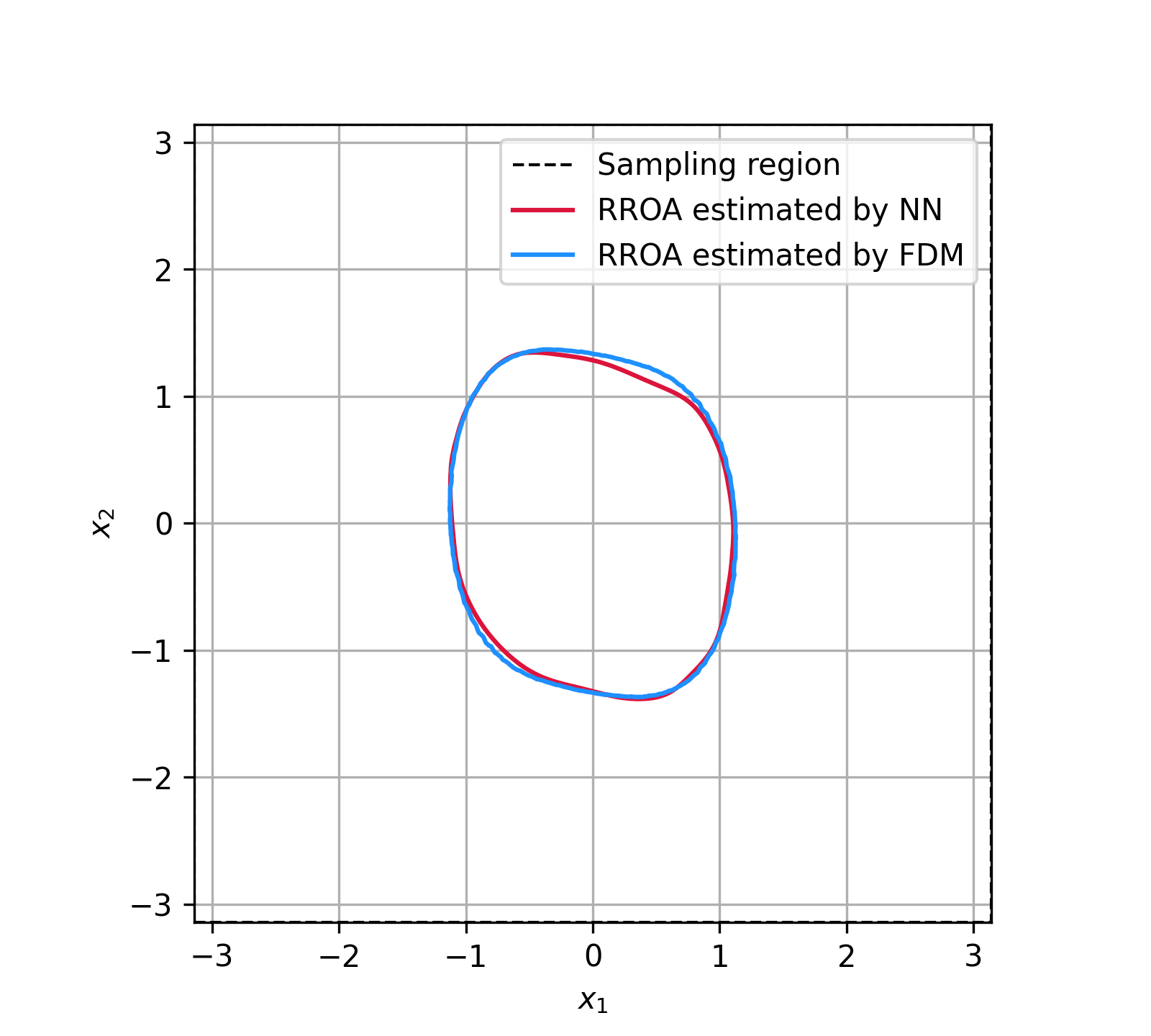}
        \caption{Learned RROA of the perturbed inverted pendulum system.}
    \label{subfig:IPRROA}
    \end{subfigure}
    \caption{The simulation result for the perturbed inverted pendulum system.}
    \label{fig:IPSystem}
\end{figure}

Another simulation we conducted is the inverted pendulum system, whose dynamics is illustrated as follows:
\begin{equation}\label{eqn:IP}
\begin{aligned}
    \dot{x}_1 &= -x_2 \\
    \dot{x}_2 &= \frac{g}{l}\sin(x_1) - \frac{b}{ml^2}x_2 + \frac{1}{ml^2}u 
\end{aligned}
\end{equation}
with the input designed as
\begin{equation}
    u = -10.2x_1-0.5x_2+\delta_1x_1 + \delta_2x_2.
\end{equation}
We chose $g = 9.81,\ l=1.0,\ b = 0.1$ and $m = 1.0$ for validation. The disturbances were chosen as $(\delta_1,\delta_2) \in [-0.3, 0.3]\times[-0.2,0.2]$. The sampling region was chosen as $\Omega=[-\pi,\pi]\times[-\pi, \pi]$. The learned value function is shown as Fig.\ref{subfig:IPValueFunction}. We also compared the RROA estimation result with FDM. As shown in Fig.\ref{subfig:IPRROA}, the neural network-based estimation result is consistent with FDM, demonstrating the accuracy of the proposed method.

\subsubsection{High-dimensional system}
\begin{figure}[h]
    \centering
    \begin{subfigure}[t]{0.45\linewidth}        \includegraphics[width=\linewidth]{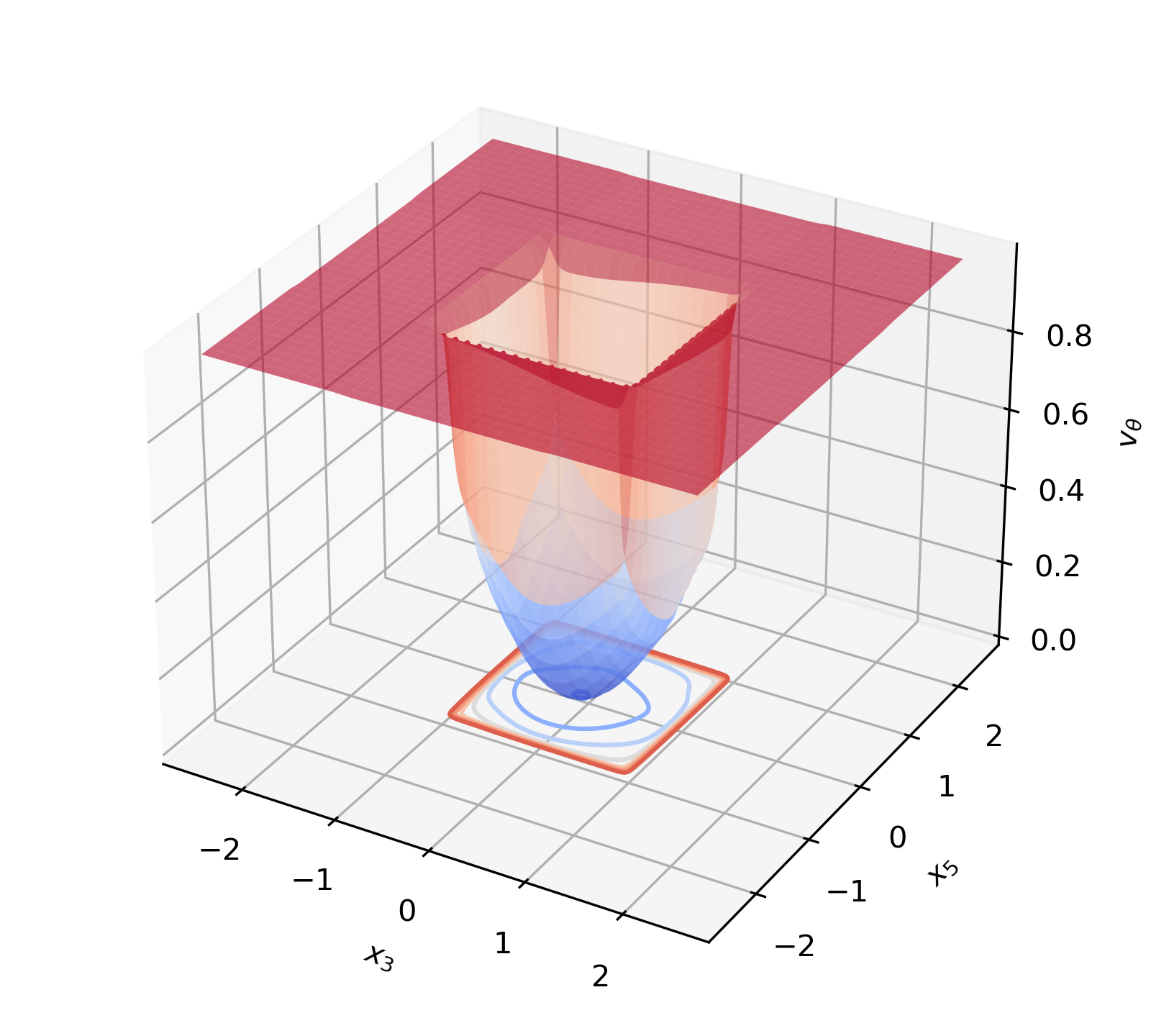}
        \caption{Learned $v_{\bm{\theta}}(\bm{x})$ of the 10-dimensional system in the $x_3-x_5$ plane with all other state components fixed at zero. }
        \label{subfig:10dValueFunction}
    \end{subfigure}
    \hfill
    \begin{subfigure}[t]
    {0.45\linewidth}
    \includegraphics[width=\linewidth]{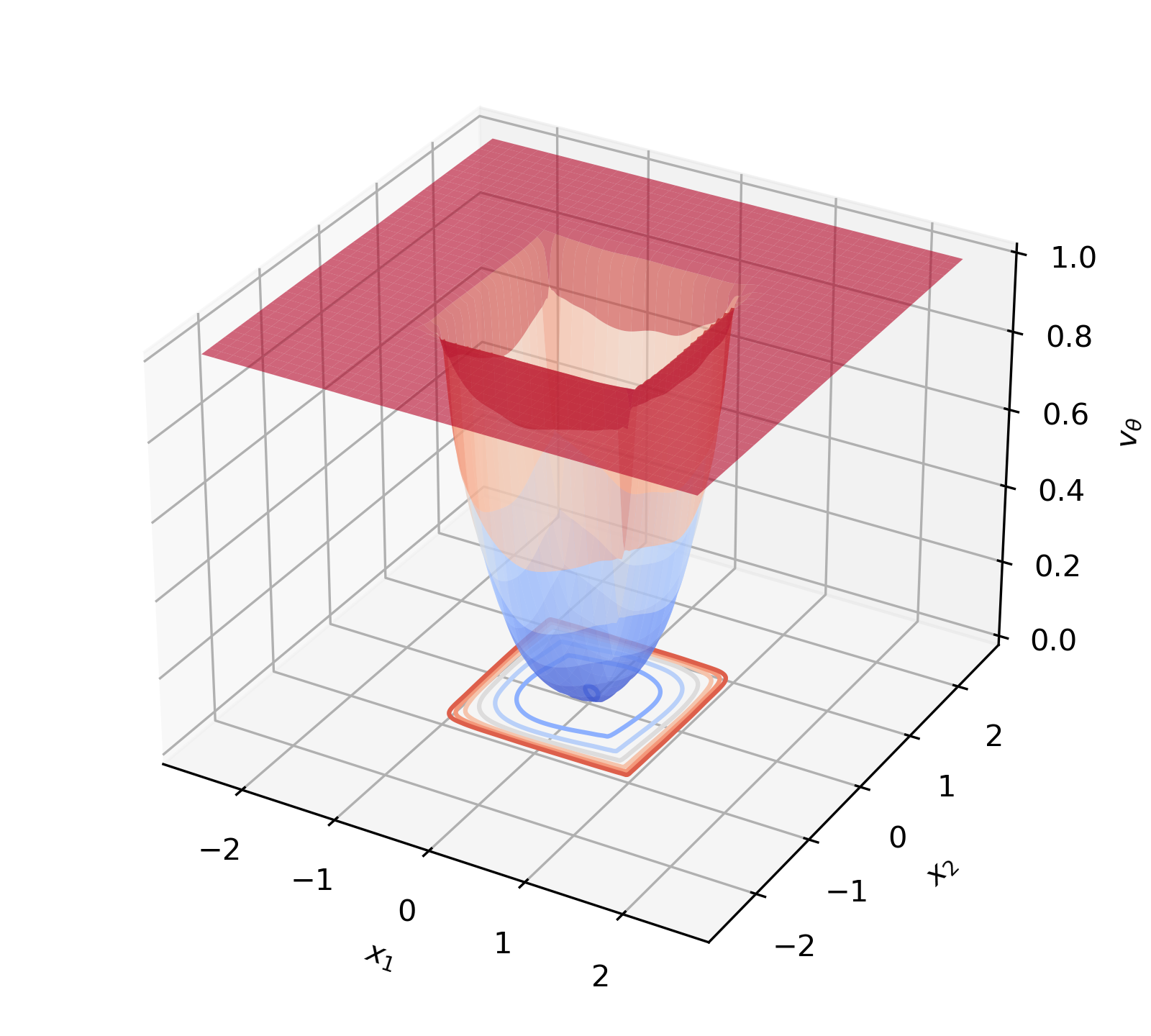}
    \caption{Learned $v_{\bm{\theta}}(\bm{x})$ of the reduced 2-dimensional system.}
    \label{subfig:2dValueFunction}
    \end{subfigure}
    \hfill
    \begin{subfigure}[t]{0.45\linewidth}
    \includegraphics[width=\linewidth]{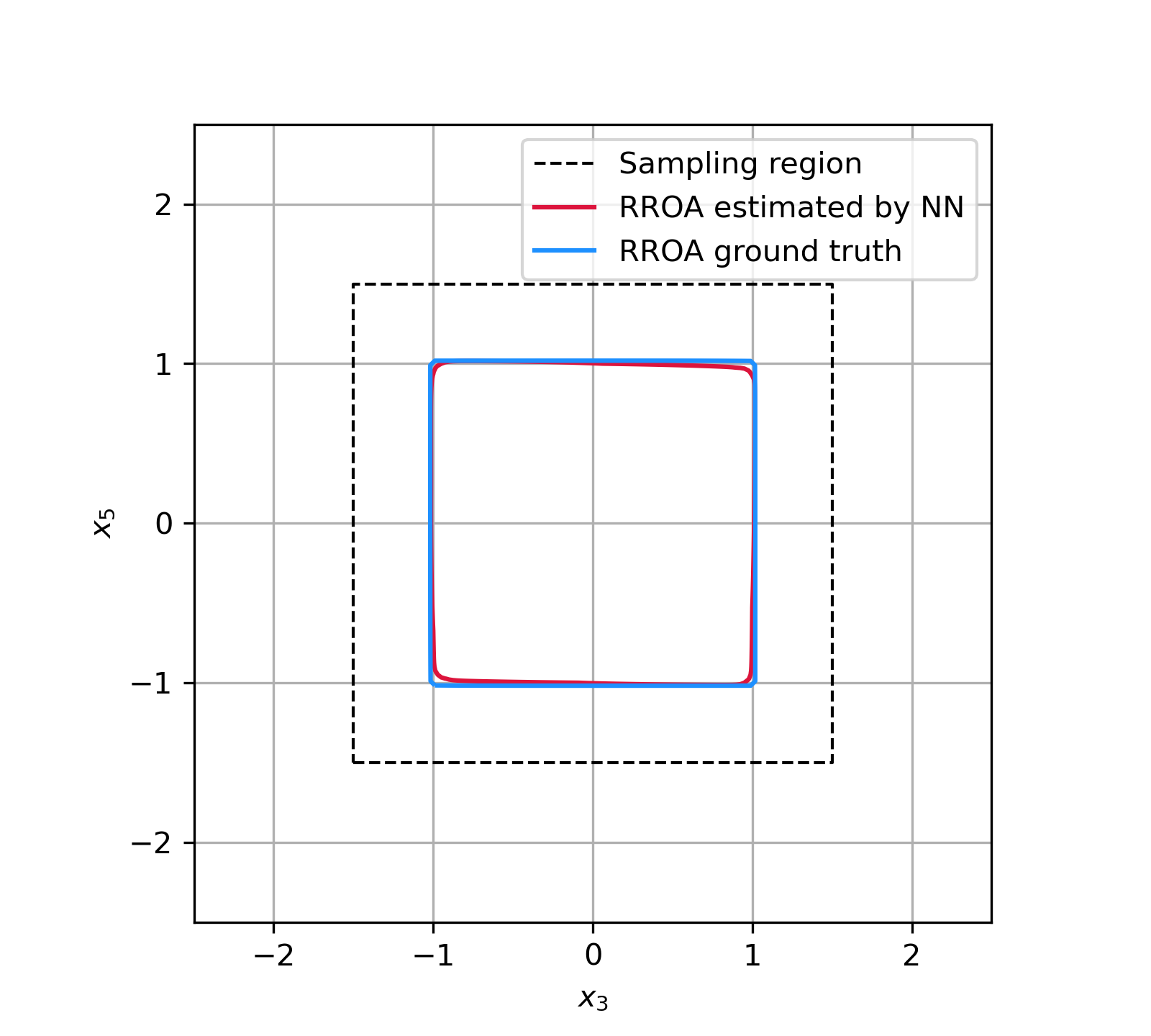}
        \caption{Learned RROA of 10-dimensional system in the $x_3-x_5$ plane with all other state components fixed at zero.}
    \label{subfig:10dRROA}
    \end{subfigure}
    \caption{The simulation result of the 10-dimensional system.}
    \label{fig:10d}
\end{figure}

To evaluate the functionality of the proposed method in high-dimensional systems, we carried out a simulation on the following system:
\begin{equation}
\begin{aligned}
    \dot{x}_i = -x_i + \delta x_i^2,
\end{aligned}
\end{equation}
where $i\in\{1,\cdots,10\}$ and $\delta\in[-1, 1]$. The sampling region was chosen as $\Omega = [-1.5,1.5]^{10}$. The ground truth RROA is $[-1, 1]^{10}$. The training results for the 10-dimensional system are shown in Fig.\ref{fig:10d}. To visualize the learned value function, we randomly select two dimensions, $x_3$ and $x_5$, while setting all other states to 0. The corresponding slice of the value function is presented in Fig.\ref{subfig:10dValueFunction}. For comparison, Fig.\ref{subfig:2dValueFunction} shows the directly learned value function of the reduced 2-dimensional system, i.e., $i=\{1,2\}$. Ideally, these two value functions should coincide. In practice, only a slight distortion is observed in the 10-dimensional case. Fig.\ref{subfig:10dRROA} shows that the learned RROA closely approximates the ground truth. These two facts demonstrate the capability of the proposed method in learning the RROA for high-dimensional systems.

% \begin{figure}[htbp]
%     \centering
%     \begin{subfigure}[t]{0.45\linewidth}        \includegraphics[width=\linewidth]{Figures/NNdynamics10d.png}
%         \caption{Learned $v_{\bm{\theta}}(\bm{x})$ for the perturbed 10-dimensional system.}
%         \label{subfig:VDPValueFunction}
%     \end{subfigure}
%     \hfill
%     \begin{subfigure}[t]{0.45\linewidth}
%     \includegraphics[width=\linewidth]{Figures/KNRdynamics10d_NNdynamics10d.png}
%         \caption{Learned RROA for the 10-dimensional system.}
%     \label{subfig:VDPRROA}
%     \end{subfigure}
%     \caption{The simulation result for the perturbed 10-dimensional system.}
%     \label{fig:VDPSystem2}
% \end{figure}

\section{Conclusions} 
\label{sec:conclusions}
In this paper, we propose a neural network framework to approximate the viscosity solution of the generalized Zubov's equation to identify the RROA for perturbed systems. Traditional grid-based numerical methods are severely limited by the curse of dimensionality. 
To overcome this challenge, we integrate a neural network with a policy iteration strategy to approximate the viscosity solution. 
In addition to computing the optimal disturbance during policy improvement, the neural network is also employed to in the rollout process to determine the optimal disturbance, thereby providing estimated anchor values to enhance training efficiency.
Simulation results demonstrate that the proposed framework effectively avoids singularities and accurately estimates the RROA in both low- and high-dimensional systems.

In future work, we plan to provide a convergence analysis of this method. Moreover, a key limitation of the current approach is its dependence on known system dynamics and perturbation bounds, which pose obstacles for this method to real-world systems. We aim to extend the method to scenarios where such information is partially or entirely unknown.

%
%%==== References ====
%
\bibliographystyle{IEEEtran}
\bibliography{reference}

\end{document}